\RequirePackage{amsmath}
\documentclass[a4paper,oribibl,envcountsame,draft,runningheads]{llncs}

\usepackage{version}

\excludeversion{rep}
\includeversion{conf}

\newcommand\maybesubsection[1]{\ourparagraph{#1}}

\def\orcid#1{{\href{http://orcid.org/#1}{\protect\raisebox{-1.25pt}{\protect\includegraphics{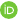}}}}}
\usepackage[final]{graphicx}
\usepackage{lmodern}
\PassOptionsToPackage{hyphens}{url}
\usepackage{iftex}
\ifpdftex
    \usepackage[T1]{fontenc}
\fi
\usepackage{marvosym}
\usepackage{amssymb}
\usepackage{bm}
\usepackage{booktabs}
\usepackage{centernot}
\usepackage{cite}
\usepackage{color}
\usepackage{ebproof}
\usepackage{enumerate}
\usepackage{stmaryrd}
\usepackage{url}
\usepackage{makecell}
\usepackage{mathtools}
\usepackage{microtype}
\usepackage{array}
\usepackage{tabu}
\usepackage{prftree}
\usepackage{float}
\usepackage{tikz-cd}
\usepackage{tikz}
\usepackage{xr}
\usepackage{xcolor}

\usepackage[mathscr]{eucal}

\usetikzlibrary{graphs, graphs.standard, quotes}

\includeversion{old-argcong}
\excludeversion{new-argcong}

\makeatletter
\g@addto@macro{\UrlBreaks}{\UrlOrds\do\=\do\_}
\makeatother

\usepackage[
    pdftitle={Exploiting Instantiations from Paramodulation Proofs in Isabelle/HOL},
    pdfauthor={Lukas Bartl, Jasmin Blanchette, and Tobias Nipkow},
    pdfkeywords={instantiations, paramodulation, Isabelle/HOL, Sledgehammer},
    pdfborder={0 0 0},
    draft=false,
    bookmarksnumbered,
    bookmarks,
    bookmarksdepth=2,
    bookmarksopenlevel=2,
    bookmarksopen]{hyperref}

\DeclareFontFamily{OT1}{pzc}{}
\DeclareFontShape{OT1}{pzc}{m}{it}{<-> s * [1.10] pzcmi7t}{}
\DeclareMathAlphabet{\mathcalx}{OT1}{pzc}{m}{it}
\let\oldlabelitemi=\labelitemi
\let\labelitemi=\labelitemii %% CHEAT!
\let\labelitemii=\oldlabelitemi %% CHEAT!

\def\negvthinspace{\kern-0.083333em}
\def\vthinspace{\kern+0.083333em}
\def\vvthinspace{\kern+0.0416667em}
\def\negvvthinspace{\kern-0.0416667em}

\newcommand\CORR{(\Letter)}

\spnewtheorem{lemmax}[theorem]{Lemma}{\bfseries}{\slshape}
\spnewtheorem{corollaryx}[theorem]{Corollary}{\bfseries}{\slshape}
\spnewtheorem{theoremx}[theorem]{Theorem}{\bfseries}{\slshape}
\spnewtheorem{conventionx}{Convention}{\bfseries}{\rmfamily}
\spnewtheorem{definitionx}{Definition}{\bfseries}{\rmfamily}
\spnewtheorem{notationx}{Notation}{\bfseries}{\rmfamily}
\spnewtheorem{examplex}{Example}{\bfseries}{\rmfamily}
\spnewtheorem{remarkx}{Remark}{\bfseries}{\rmfamily}

\let\oldSigma=\Sigma
\renewcommand\Sigma{\mathrm{\oldSigma}}

\newcommand\medrightarrow{\mathrel{{{\color{black}\relbar}\kern-0.9ex\rlap{\color{white}\ensuremath{\blacksquare}}\kern-0.9ex}\joinrel{\color{black}\rightarrow}}}
\newcommand\medleftarrow{\mathrel{{\color{black}\leftarrow}\kern-0.9ex\rlap{\color{white}\ensuremath{\blacksquare}}\kern-0.9ex\joinrel{{\color{black}\relbar}}}}
\newcommand\medleftrightarrow{\mathrel{\leftarrow\kern-1.685ex\rightarrow}}
\newcommand\Medrightarrow{\mathrel{{{\color{black}\Relbar}\kern-0.9ex\rlap{\color{white}\ensuremath{\blacksquare}}\kern-0.9ex}\joinrel{\color{black}\Rightarrow}}}
\newcommand\Medleftrightarrow{\mathrel{\Leftarrow\kern-1.685ex\Rightarrow}}

\newcommand\ourparagraph[1]{\paragraph{\bfseries\upshape#1.}}

\begin{document}

\begin{conf}\title{Exploiting Instantiations from Paramodulation~Proofs in Isabelle/HOL}\end{conf}
\begin{rep}\title{Exploiting Instantiations from Paramodulation~Proofs in Isabelle/HOL (Technical~Report)}\end{rep}
\titlerunning{Exploiting Instantiations from Paramodulation Proofs in Isabelle/HOL}
\author{Lukas Bartl\inst{1}\textsuperscript{\CORR}\orcid{0009-0000-2439-5025}
  \and
  Jasmin Blanchette\inst{2}\orcid{0000-0002-8367-0936}
  \and
  Tobias Nipkow\inst{3}\orcid{0000-0003-0730-515X}
  }
\authorrunning{L. Bartl et al.}
% First names are abbreviated in the running head.
% If there are more than two authors, 'et al.' is used.
%
\institute{Universität Augsburg, Augsburg, Germany \\
\email{lukas.bartl@uni-a.de}
\and
Ludwig-Maximilians-Universität München, Munich, Germany \\
\email{jasmin.blanchette@ifi.lmu.de}
\and
Technische Universität München, Munich, Germany \\
\email{nipkow@in.tum.de}}

\maketitle

\begin{abstract}
Metis is an ordered paramodulation prover built into the Isabelle/HOL
proof assistant. It attempts to close the current goal using a given list of
lemmas. Typically these lemmas are found by Sledgehammer, a tool that integrates
external automatic provers. We present a new tool that analyzes successful Metis
proofs to derive variable instantiations.
These increase Sledgehammer's success rate,
improve the speed of Sledgehammer-generated proofs,
and
help users understand why a goal follows from the lemmas.
\end{abstract}

\section{Introduction}
\label{sec:introduction}

Sledgehammer~\cite{Paulson2012} is undoubtedly one of the interactive proof
assistant Isabelle/HOL's \cite{Nipkow2002} most popular components. This
component gives Isabelle access to many external automatic theorem provers (ATPs),
which can mechanize routine proofs \cite{boehme-nipkow-2010,Desharnais2022}.
When the user
invokes Sledgehammer on a proposition to be proved, it performs the
following steps:
\begin{enumerate}
\item It invokes a number of external ATPs
  with the goal and the background library of \emph{facts}
  (definitions, lemmas, theorems, etc.);
  this involves translating the goal and facts from higher-order logic to the
  ATPs' logics.%
\smallskip
\item For any proof that is found, the used facts are extracted from it
 and the internal prover
  \emph{Metis}~\cite{Hurd2003}, based on ordered paramodulation, is
  invoked to reconstruct an Isabelle proof based on this list of facts
  only.
\end{enumerate}

This paper is about obtaining and exploiting the instantiations of the
facts used in the proofs. Our motivation is that using instantiated facts
\begin{samepage}
    \begin{itemize}
        \item can speed up proof reconstruction (and even turn timeouts into successes);%
        \smallskip
        \item can improve readability of proofs for humans; and%
        \smallskip
        \item can lead to simpler proofs.
    \end{itemize}
\end{samepage}
\begin{examplex}
We start with a simple example where we believe human readability
is improved.  We work in a linear order with $\bot$ and $\top$.
Sledgehammer proves that in the context of the facts
\[\top \nless x \enspace (1) \qquad x \neq \bot \leftrightarrow \bot < x \enspace (2)\]
the assumption $\bot \nless \top$ implies $\mathtt{a} = \mathtt{b}$,
where $\mathtt{a}$ and $\mathtt{b}$ are arbitrary constants.
Why is that?  It turns out that the following instantiations are used:
\begin{gather*}
    \top \nless \mathtt{a} \enspace (1\mathtt{a}) \qquad
    \top \nless \mathtt{b} \enspace (1\mathtt{b}) \\
    \mathtt{a} \neq \bot \leftrightarrow \bot < \mathtt{a} \enspace (2\mathtt{a}) \qquad
    \mathtt{b} \neq \bot \leftrightarrow \bot < \mathtt{b} \enspace (2\mathtt{b}) \qquad
    \top \neq \bot \leftrightarrow \bot < \top \enspace (2\top)
\end{gather*}
Clearly $\bot \nless \top$ implies $\bot=\top$ via $(2\top)$. This
implies $\bot \nless \mathtt{a}$ and $\bot \nless \mathtt{b}$ via $(1\mathtt{a})$ and $(1\mathtt{b})$. Now we
obtain $\mathtt{a} = \bot$ and $\mathtt{b} = \bot$ via $(2\mathtt{a})$ and $(2\mathtt{b})$, and thus $\mathtt{a}=\mathtt{b}$.
Although this deduction still needs some thought, we believe
that it is far simpler than if we had to work out the derivation from $(1)$ and $(2)$ alone.
\end{examplex}

This paper is structured as follows.
Section~\ref{sec:background} presents the necessary background about
Isabelle, Sledgehammer, and Metis. Section~\ref{sec:examples}
presents further introductory examples. The core of the paper is in
Section~\ref{sec:metis-extension}, which explains how to obtain the
instantiations used during a Metis proof and translate them into
instantiations of the corresponding Isabelle
facts. Section~\ref{sec:sh-extension} explains how Sledgehammer was
extended to cooperate with the extended Metis
prover. Section~\ref{sec:evaluation} presents our empirical
evaluation of how instantiations improve the performance of
Sledgehammer.
We end with related work and the conclusion.

Our extensions are available as part of Isabelle starting with version 2025.%
\footnote{\url{https://isabelle.in.tum.de/website-Isabelle2025/}}
They are documented in the Sledgehammer user's manual \cite{sledgehammer-manual}.
The raw data for our evaluation is available online.%
\footnote{\url{https://nekoka-project.github.io/pubs/instantiations_data.zip}}
%\begin{conf}%
%A proof of our metatheorem is included in a technical report \cite{our-report}.
%\end{conf}%

\section{Background}
\label{sec:background}

\begin{rep}
Our work is based on Isabelle (Section~\ref{ssec:isabelle}). Particularly relevant
are its Sledgehammer advisory tool (Section~\ref{ssec:sledgehammer}) and
its \emph{metis} proof method (Section~\ref{ssec:the-metis-proof-method}),
which relies on the Metis ATP (Section~\ref{ssec:the-metis-atp}).
\end{rep}

\maybesubsection{Isabelle}
\label{ssec:isabelle}
Isabelle/HOL \cite{Wenzel-PhD,Nipkow2002} is a proof assistant for polymorphic
higher-order logic enriched with type classes \cite{RosskopfN-JAR23}. It is
written primarily in Standard ML. It has an inference kernel through which all
logical inferences must go to be deemed acceptable. At the user level,
notations largely follow mathematical practice.

\maybesubsection{Sledgehammer}
\label{ssec:sledgehammer}

The Sledgehammer tool \cite{Paulson2012} consists of six main
components:
\begin{enumerate}
\item The \emph{relevance filter} (or ``premise selector'') heuristically
  selects a subset of the available facts as likely relevant to the current
  goal. Typically, hundreds of facts can be chosen without overwhelming ATPs.
  Sledgehammer includes two relevance filters
  \cite{meng-paulson-2009-relev,blanchette-et-al-2016-mash}, which can be
  combined.%
\smallskip
\item The \emph{translation module} constructs an ATP problem from the
  selected facts and the current goal, translating Isabelle's polymorphic
  higher-order logic to the ATP's logic \cite{Meng2007,Blanchette2016}.%
\smallskip
\item The {ATP} tries to prove the problem. Actually, multiple ATPs can be
  run in parallel.
  Commonly used ATPs include
  cvc5 \cite{barbosa-et-al-2022}, E \cite{vukmirovic-et-al-2023}, Leo-III
  \cite{steen-benzmueller-2018}, SPASS \cite{blanchette-et-al-2012-spass},
  Vampire \cite{bhayat-suda-2024}, veriT \cite{bouton-et-al-2009},
  Z3 \cite{MouraB08},
  and
  Zipperposition \cite{vukmirovic-et-al-2021-making}.%
\smallskip
\item If one or more ATPs find a proof, the \emph{proof minimization module}
  repeatedly invokes each ATP with subsets of the facts referenced in the
  respective proof,
  trying to reduce the number of dependencies and speed up the next steps.%
\smallskip
\item The \emph{proof reconstruction module} transforms each ATP proof
  into a textual Isabelle proof. Reconstruction means that
  Sledgehammer and the ATP need not be trusted. Typically, the structure
  of the ATP proof is discarded, and the Isabelle proof consists of a single
  proof method (often \textit{metis}) invoked with the facts referenced in the ATP proof. Detailed
  Isabelle proofs, or Isar proofs \cite{Wenzel99}, are available as an
  experimental feature \cite{blanchette-et-al-2016-isar}.%
\smallskip
\item For each ATP proof, the \emph{preplay module} tries out various proof
  methods before they are presented to the user. If several methods
  succeed, the fastest one is chosen \cite{blanchette-et-al-2016-isar}.
\end{enumerate}

%\begin{rep}
As an example, suppose Sledgehammer selects 512 facts $f_1, \dots, f_{512}$ and
passes them, along with the goal, to E. Then E finds a proof involving three
facts, $f_{10}$, $f_{73}$, and $f_{359}$, and minimization reduces this list to
two:\ $f_{10}$ and $f_{359}$. By trial and error (and preplaying),
Sledgehammer determines that the proof method \textit{metis} with $f_{10}$ and
$f_{359}$ as arguments solves the Isabelle goal in \(23\)\,ms, and no other proof
method succeeds, so this \textit{metis} call is suggested to the user.
%\end{rep}

\maybesubsection{The Metis ATP}
\label{ssec:the-metis-atp}

Metis \cite{Hurd2003} is an ATP for untyped first-order logic with equality
written in Standard ML. It is based on ordered paramodulation, a variant of
superposition \cite{bachmair-ganzinger-1990-on}. Although Metis was developed
as a standalone program, it is also incorporated in Isabelle's source code, so
that it is always available.

Thanks to its calculus, Metis is reasonably performant, although it cannot
compete with state-of-the-art superposition provers such as Vampire
\cite{DBLP:journals/aicom/Sutcliffe12}. Metis's main strengths are the
readability of its source code and the simplicity and fine granularity of its
proof format. Proofs are expressed using the following six inference rules:

\vskip\abovedisplayskip

\noindent\hbox{}\hfill
\begin{prooftree}
\hypo{\phantom{C}}
\infer1[\textsc{Axiom}]
{C}
\end{prooftree}%
\hfill
\begin{prooftree}
\hypo{\phantom{C}}
\infer1[\textsc{Assume}]
{A \lor \lnot A}
\end{prooftree}%
\hfill
\begin{prooftree}
\hypo{C}
\infer1[\textsc{Subst}]
{C\sigma}
\end{prooftree}%
\hfill\hbox{}

\medskip

\noindent\hbox{}\hfill
\begin{prooftree}
\hypo{\phantom{C}}
\infer1[\textsc{Refl}]
{t = t}
\end{prooftree}%
\hfill\hfill
\begin{prooftree}
\hypo{\phantom{C}}
\infer1[\textsc{Equality}]
{s \neq t \lor \lnot L[s]_p \lor L[t]_p}
\end{prooftree}%
\hfill\hfill
\begin{prooftree}
\hypo{C \lor A}
\hypo{\lnot A \lor D}
\infer2[\textsc{Resolve}]
{C \lor D}
\end{prooftree}%
\hfill\hbox{}

\vskip\belowdisplayskip

Notice that substitution is captured by the explicit \textsc{Subst} rule instead of being
part of \textsc{Resolve}. Moreover, all equality reasoning is reduced to
\textsc{Refl} and \textsc{Equality}. For proof search, Metis relies on a more
efficient calculus that performs ordered paramodulation, but the proofs are then
translated to the above fine-granular rules.

\maybesubsection{The \textit{metis} Proof Method}
\label{ssec:the-metis-proof-method}

Isabelle's \emph{metis} (with a lowercase~m) proof method \cite{Paulson2007}
builds on the Metis (with an uppercase M) ATP to provide general-purpose proof
automation. The \emph{metis} proof method takes a list of facts as argument
and translates them, together with the current goal, from polymorphic
higher-order logic to untyped first-order logic.
The resulting \emph{axiom clauses} are then introduced in the proof attempt
using the \textsc{Axiom} rule. The translation uses the same
techniques (and the same code) as Sledgehammer. Next, \emph{metis} invokes
the Metis ATP, and if Metis finds a proof, it is reconstructed step by
step using Isabelle's inference kernel, so that the goal becomes an Isabelle
theorem. A \textit{metis} call is considered successful if the Metis ATP found
a proof and Isabelle reconstructed~it.

Since it may be hard for the user to determine which facts are
necessary for a proof, in practice \emph{metis} is almost always used in
conjunction with Sledgehammer.

\section{Examples}
\label{sec:examples}

Before we study our \textit{metis} and Sledgehammer extensions in detail,
we take a look at a few examples that illustrate how instantiating facts can not only improve readability but also
speed up proof reconstruction and lead to simpler proofs.  The final example demonstrates how more
complex terms are displayed.

\begin{examplex}
\label{ex:timeout}
    We present a goal from the \emph{Archive of Formal Proofs} \cite{Afp2015}, where
    instantiating facts substantially speeds up the proof and turns a timeout into a success.  This
    example stems from a formalization of Tarski's axioms for Euclidean geometry
    \cite{IsaGeoCoq-AFP}.  The lemma \textit{cong\_mid2\_\_cong} states the following property,
    where \(\mathtt{Cong}\) denotes the congruence relation and \(\mathtt{Midpoint}\) states that a
    point lies exactly in the middle of two other points:
    \[\mathtt{Midpoint}\;\mathtt{M}\;\mathtt{A}\;\mathtt{B} \longrightarrow
    \mathtt{Midpoint}\;\mathtt{M}'\;\mathtt{A}'\;\mathtt{B}' \longrightarrow
    \mathtt{Cong}\;\mathtt{A}\;\mathtt{M}\;\mathtt{A}'\;\mathtt{M}' \longrightarrow
    \mathtt{Cong}\;\mathtt{A}\;\mathtt{B}\;\mathtt{A}'\;\mathtt{B}'\] When invoking Sledgehammer on
    a modern laptop for this goal, the external ATP cvc5 \cite{barbosa-et-al-2022} finds a proof
    using the facts \textit{cong\_inner\_transitivity}, \textit{l2\_l1\_b}, \textit{midpoint\_bet},
    and \textit{midpoint\_cong}.  However, \textit{metis} times out; i.e., it fails
    to find a proof derived from these facts within \(1\)\,s.
    After inferring the instantiations and instantiating the facts, though,
    \textit{metis} successfully solves the goal in \(50\)\,ms.  The instantiations are simple: Each
    free variable in the facts is replaced by one of \(\mathtt{M}, \mathtt{A}, \mathtt{B},
    \mathtt{M}', \mathtt{A}', \mathtt{B}'\).  The facts \textit{midpoint\_bet} and
    \textit{midpoint\_cong} each have two different instantiations and thus appear twice in the
    resulting Isabelle proof, which may also improve readability.
\end{examplex}

\begin{examplex}
\label{ex:speedup-auto}
    Instantiating facts not only speeds up \textit{metis} but also leads to
    simpler and more readable proofs.  Suppose that we want to prove the goal \(0 < \mathtt{i}
    \longrightarrow \mathtt{x} \leq \mathtt{x} ^ \mathtt{i}\) for natural numbers \(\mathtt{i},
    \mathtt{x}\).  When invoking Sledgehammer on a modern laptop, the external ATP
    veriT \cite{bouton-et-al-2009} finds a proof including the following facts, where
    \(\mathtt{Suc}\) returns the successor of its argument:
    \pagebreak[0]
    \begin{align*}
        1 = \mathtt{Suc}\;0
        &\enspace (\textit{nat\_1}) \\
        a \leq 0 \longrightarrow a = 0
        &\enspace (\textit{bot\_nat\_0.extremum\_uniqueI}) \\
        m \nleq n \longleftrightarrow \mathtt{Suc}\;n \leq m
        &\enspace (\textit{not\_less\_eq\_eq}) \\
        1 \leq a \longrightarrow 0 < n \longrightarrow a \leq a ^ n
        &\enspace (\textit{self\_le\_power}) \\
        0 \leq a \longrightarrow 0 \leq a ^ n
        &\enspace (\textit{zero\_le\_power})
    \end{align*}
    The \textit{metis} method can solve the goal with these facts in \(82\)\,ms.  However,
    after inferring the instantiations and instantiating the facts, \textit{metis} needs
    only \(41\)\,ms to solve the goal, and the \textit{auto} \cite{Paulson1994} proof method is even
    faster, needing only \(7\)\,ms.  Additionally, \textit{auto} requires only the following two
    instantiated facts:
    \begin{align*}
        \mathtt{x} \nleq 0 \longleftrightarrow \mathtt{Suc}\;0 \leq \mathtt{x}
        &\enspace
        (\textit{not\_less\_eq\_eq} \text{ with } \{m \mapsto \mathtt{x}{,}\; n \mapsto 0\}) \\
        1 \leq \mathtt{x} \longrightarrow 0 < \mathtt{i} \longrightarrow \mathtt{x} \leq \mathtt{x}
        ^ \mathtt{i}
        &\enspace (\textit{self\_le\_power} \text{ with } \{a \mapsto \mathtt{x}{,}\; n
        \mapsto \mathtt{i}\})
    \end{align*}
    By contrast, \textit{auto} fails to find a proof derived from these facts without
    instantiations.  The Isabelle proof using \textit{auto} is simpler and more readable than the \textit{metis} proof, since trivial facts were eliminated.  This helps the user focus on
    the relevant facts when trying to understand the proof.
\end{examplex}

\begin{examplex}
\label{ex:skolem-lambda}
    In the preceding examples, the instantiations were simple, with each free variable being
    replaced by a single symbol.  There are also proofs where instantiations are more
    complex, possibly involving \(\lambda\)-abstractions and quantified variables.
    Consider the goal
    \[\mathtt{surj}\;(\lambda n .\; \mathtt{g}\;(\mathtt{Suc}\;n)) \longrightarrow (\exists m .\;
    \mathtt{P}\;(\mathtt{Suc}\;(\mathtt{g}\;m))) \longrightarrow (\exists n .\;
    \mathtt{P}\;(\mathtt{g}\;(\mathtt{Suc}\;n)))\]
    where \(\mathtt{surj}\) states that a function is surjective.
    We have the fact \(\mathtt{surj}\;f \longrightarrow (\exists x .\; f\;x = y)\) at our disposal.
    Our \textit{metis} extension suggests instantiating \(f\)
    with \(\lambda c .\; \mathtt{g}\;(\mathtt{Suc}\;c)\) and \(y\) with
    \(\mathtt{Suc}\;(\mathtt{g}\;\_)\) in the fact.
    Notice that bound variables can be renamed in the inferred instantiations
    (here, \(n\) became \(c\)).
    Additionally, `\(\_\)'---which corresponds to a fresh free variable---can appear as a
    placeholder for quantified variables (e.g., \(m\)), since there is no way to refer to them.
\end{examplex}

\section{Instantiations from Metis Proofs}
\label{sec:metis-extension}

We extend the \textit{metis} proof method so that it infers the instantiations of facts used
in a proof and suggests them to the user.  This extension is executed after a successful
\textit{metis} proof, provided that the \textit{metis\_instantiate} option is enabled.

\maybesubsection{Inference of Instantiations for Metis Proofs}

To infer the instantiations of the Isabelle facts, we first infer the
instantiations of the Metis clauses. These are the substitutions that
were applied to the axiom clauses in the Metis proof.

\begin{definitionx}
    A \emph{Metis theorem} \(\theta\) is recursively defined as a triple \((C, r, \bar \theta)\) consisting of a Metis
    clause \(C\), an inference rule \(r\), and a list of Metis theorems \(\bar \theta\), where \(C\)
    is derived directly from \(\bar \theta\) using the rule \(r\).  A \emph{Metis proof} is a Metis
    theorem for the empty clause, denoted by \(\mathtt{False}\).
\end{definitionx}

\begin{definitionx}
    A \emph{Metis substitution} \(\sigma = \{x_1 \mapsto t_1, \dots, x_n \mapsto t_n\}\) is a
    partial function from Metis variable names to Metis terms.  The \emph{composition} of two
    substitutions \(\sigma_1 \circ \sigma_2\) first applies \(\sigma_2\) and then \(\sigma_1\), so
    that \(C (\sigma_1 \circ \sigma_2) = (C \sigma_2) \sigma_1\).
\end{definitionx}

\begin{definitionx}
    The function \(\operatorname{infer}(\theta, \sigma)\)
    returns a list of pairs \((C, \sigma')\), where \(C\) is an axiom
    clause in the Metis theorem \(\theta\) and \(\sigma'\) is the applied
    substitution.
    The argument \(\sigma\) is used as an accumulator and is
    initialized to \(\varnothing\) for a Metis proof.
    The function is defined recursively as follows:
    \begin{align*}
        \operatorname{infer}((C, r, []), \sigma) &=
        \begin{cases}
            [(C, \sigma)] &\text{if } r = \textsc{Axiom} \\
            [] &\text{otherwise}
        \end{cases} \\
        \operatorname{infer}((C, \textsc{Subst}, [\theta]), \sigma) &=
        \operatorname{infer}(\theta, \sigma \circ \sigma') \\
        \operatorname{infer}((C, \textsc{Resolve}, [\theta_1, \theta_2]), \sigma) &=
        \operatorname{infer}(\theta_1, \sigma) \mathbin{@} \operatorname{infer}(\theta_2, \sigma)
    \end{align*}
    where $\sigma'$ in the second equation is the substitution carried by the
    \textsc{Subst} inference rule on the left, and \(@\) in the third equation
    denotes list concatenation.
\end{definitionx}

\begin{examplex}
    Suppose that we want to prove the goal \(1 < \mathtt{Suc}\;(\mathtt{Suc}\;\mathtt{x})\)
    on natural numbers from the following facts:
    \[m < n \longrightarrow \mathtt{Suc}\;m < \mathtt{Suc}\;n \qquad \mathtt{Suc}\;0 = 1 \qquad 0 <
    \mathtt{Suc}\;n\]
    We translate the goal and the facts into first-order logic by using the predicate symbol
    \(\mathtt{less}\) for \(<\).  Metis finds the following proof:
    \begin{enumerate}[(1)]
        \item% 1
            \textsc{Axiom}: \(\lnot \mathtt{less}(1, \mathtt{Suc}(\mathtt{Suc}(\mathtt{x})))\)
        \smallskip
        \item% 2
            \textsc{Axiom}: \(\lnot \mathtt{less}(m, n) \lor \mathtt{less}(\mathtt{Suc}(m),
            \mathtt{Suc}(n))\)
        \smallskip
        \item% 3
            \textsc{Subst} from (2) using \(\sigma = \{m \mapsto 0{,}\; n \mapsto y\}\):\\ \(\lnot
            \mathtt{less}(0, y) \lor \mathtt{less}(\mathtt{Suc}(0), \mathtt{Suc}(y))\)
        \smallskip
        \item% 4
            \textsc{Axiom}: \(\mathtt{Suc}(0) = 1\)
        \smallskip
        \item% 5
            \textsc{Equality}: \(\mathtt{Suc}(0) \neq 1 \lor \lnot
            \mathtt{less}(\mathtt{Suc}(0), \mathtt{Suc}(y)) \lor \mathtt{less}(1,
            \mathtt{Suc}(y))\)
        \smallskip
        \item% 6
            \textsc{Resolve} from (4) and (5): \(\lnot \mathtt{less}(\mathtt{Suc}(0),
            \mathtt{Suc}(y)) \lor \mathtt{less}(1, \mathtt{Suc}(y))\)
        \smallskip
        \item% 7
            \textsc{Resolve} from (3) and (6): \(\lnot \mathtt{less}(0, y) \lor \mathtt{less}(1,
            \mathtt{Suc}(y))\)
        \smallskip
        \item% 8
            \textsc{Subst} from (7) using \(\sigma = \{y \mapsto \mathtt{Suc}(\mathtt{x})\}\):\\
            \(\lnot \mathtt{less}(0, \mathtt{Suc}(\mathtt{x})) \lor \mathtt{less}(1,
            \mathtt{Suc}(\mathtt{Suc}(\mathtt{x})))\)
        \smallskip
        \item% 9
            \textsc{Resolve} from (1) and (8): \(\lnot \mathtt{less}(0, \mathtt{Suc}(\mathtt{x}))\)
        \smallskip
        \item% 10
            \textsc{Axiom}: \(\mathtt{less}(0, \mathtt{Suc}(n))\)
        \smallskip
        \item% 11
            \textsc{Subst} from (10) using \(\sigma = \{n \mapsto \mathtt{x}\}\):
            \(\mathtt{less}(0, \mathtt{Suc}(\mathtt{x}))\)
        \smallskip
        \item% 12
            \textsc{Resolve} from (9) and (11): \(\mathtt{False}\)
    \end{enumerate}
    We can now infer the substitutions that were applied to the axiom clauses:
    \begin{align*}
        \operatorname{infer}(12, \varnothing) &= \operatorname{infer}(1, \varnothing) \mathbin{@}
        \operatorname{infer}(2, \{y \mapsto \mathtt{Suc}(\mathtt{x})\} \circ \{m \mapsto 0{,}\; n
        \mapsto y\}) \\[-\jot] &\phantom{{}={}} \mathbin{@} \operatorname{infer}(4, \{y \mapsto
        \mathtt{Suc}(\mathtt{x})\}) \mathbin{@} \operatorname{infer}(5, \{y \mapsto
        \mathtt{Suc}(\mathtt{x})\}) \\[-\jot] &\phantom{{}={}} \mathbin{@} \operatorname{infer}(10,
        \{n \mapsto \mathtt{x}\}) \\ &= \operatorname{infer}(1, \varnothing) \mathbin{@}
        \operatorname{infer}(2, \{y \mapsto \mathtt{Suc}(\mathtt{x}){,}\; m \mapsto 0{,}\; n \mapsto
        \mathtt{Suc}(\mathtt{x})\}), \\[-\jot] &\phantom{{}={}} \mathbin{@} \operatorname{infer}(4,
        \{y \mapsto \mathtt{Suc}(\mathtt{x})\}) \mathbin{@} [] \mathbin{@} \operatorname{infer}(10,
        \{n \mapsto \mathtt{x}\}) \\ &= [(\lnot \mathtt{less}(1,
        \mathtt{Suc}(\mathtt{Suc}(\mathtt{x}))), \varnothing), \\[-\jot] &\phantom{{}={}[} (\lnot
        \mathtt{less}(m, n) \lor \mathtt{less}(\mathtt{Suc}(m), \mathtt{Suc}(n)), \\[-\jot]
        &\phantom{{}={}[(} \{y \mapsto \mathtt{Suc}(\mathtt{x}){,}\; m \mapsto 0{,}\; n \mapsto
        \mathtt{Suc}(\mathtt{x})\}), \\[-\jot] &\phantom{{}={}[} (\mathtt{Suc}(0) = 1, \{y \mapsto
        \mathtt{Suc}(\mathtt{x})\}), (\mathtt{less}(0, \mathtt{Suc}(n)), \{n \mapsto \mathtt{x}\})]
    \end{align*}
\end{examplex}

Given a Metis proof \(\theta\), the call \(\operatorname{infer}(\theta, \varnothing)\) yields a list
of pairs \((C, \sigma)\), where \(C\) is an axiom clause and \(\sigma\) is the applied substitution.
If the proof is retried, the \emph{instantiated axiom clauses}---i.e., the list
of all clauses \(C \sigma\) corresponding to the pairs \((C, \sigma)\)---can be used instead of the
original axiom clauses to restrict the search space and speed up Metis.  After instantiation, a
Metis proof is still possible:

\begin{theoremx}
\label{thm:instantiated-axiom-clauses}
    A Metis proof\/ \(\theta\) can be transformed into a new Metis proof derived from the
    instantiated axiom clauses (defined above). The new Metis proof does not involve the
    \textnormal{\textsc{Subst}} inference rule and uses at most as many proof steps as\/ \(\theta\).
\end{theoremx}
The proof is by induction on the structure of the Metis proof \(\theta\).
It proceeds by instantiating the clauses in \(\theta\) with
\(\operatorname{infer}\)'s current accumulator
value and removing \textsc{Subst} inference steps.
Because clauses are sets and not multisets of literals,
instantiation can shorten clauses by unifying literals.
This can result in the elimination of resolution steps and thereby entire subproofs.
%\begin{rep}
\begin{proof}
    We annotate each Metis theorem in \(\theta\) with \(\operatorname{infer}\)'s current accumulator
    value using the call \(\operatorname{annotate}(\theta, \varnothing)\), defined as follows:
    \pagebreak[0]
    \begin{align*}
        \operatorname{annotate}((C, r, []), \sigma) &= (C, r, [], \sigma) \\
        \operatorname{annotate}((C, \textsc{Subst}, [\theta']), \sigma) &=
        (C, \textsc{Subst}, [\operatorname{annotate}(\theta', \sigma \circ \sigma')], \sigma) \\
        \operatorname{annotate}((C, \textsc{Resolve}, [\theta_1, \theta_2]), \sigma) &=
        (C, \textsc{Resolve}, \\[-\jot]
        &\phantom{{} = {} (} [\operatorname{annotate}(\theta_1, \sigma),
        \operatorname{annotate}(\theta_2, \sigma)], \sigma)
    \end{align*}
    where \(\sigma'\) is again the substitution carried by the \textsc{Subst} inference rule.

    It suffices to show that each annotated Metis theorem \((C, r, \bar \theta, \sigma)\)
    contained in \(\operatorname{annotate}(\theta, \varnothing)\) can be transformed into a new
    Metis theorem \((C', r', \bar \theta')\) such that \(C' \subseteq C \sigma\) holds (with \(C' =
    C \sigma\) for axiom clauses), \textsc{Subst} inference steps are removed, and no new proof step
    is introduced. Since \(C' \subseteq \mathtt{False}\, \varnothing\) implies \(C' =
    \mathtt{False}\), the whole Metis proof \(\theta\) is transformed into a new Metis proof.  By
    the definitions of the \(\operatorname{infer}\) and \(\operatorname{annotate}\) functions, \(C'
    = C \sigma\) for axiom clauses implies that the new Metis proof is derived from the instantiated
    axiom clauses.

    Assume that we have an annotated Metis theorem \((C, r, \bar \theta, \sigma)\) contained in
    \(\operatorname{annotate}(\theta, \varnothing)\).  We prove the proposition by induction on the
    inference rule \(r\).  In the base case, \(r\) has no premises, so \(\bar \theta = []\) and
    \(r\) is one of \textsc{Axiom}, \textsc{Assume}, \textsc{Refl}, and \textsc{Equality}.  We
    construct the new Metis theorem \((C', r', \bar \theta') = (C \sigma, r, [])\) so that \(C' = C
    \sigma \subseteq C \sigma\) holds.

    If \(r = \textsc{Subst}\), we have \(\bar \theta = [\theta_1]\), where \(\theta_1 = (C_1, r_1,
    \bar \theta_1)\), and \(C = C_1 \sigma'\), where \(\sigma'\) is the substitution carried by the
    \textsc{Subst} inference rule.  Inductively, we can transform
    \(\operatorname{annotate}(\theta_1, \sigma \circ \sigma')\) into a new Metis theorem \(\theta' =
    (C', r', \bar \theta')\) with \(C' \subseteq C_1 (\sigma \circ \sigma')\).  Using \(C_1 (\sigma
    \circ \sigma') = (C_1 \sigma') \sigma = C \sigma\), we can deduce \(C' \subseteq C \sigma\), so
    we use \(\theta'\) as the new Metis theorem and remove the substitution step.

    If \(r = \textsc{Resolve}\), then \(\bar \theta = [\theta_1, \theta_2]\), where \(\theta_1 =
    (C_1, r_1, \bar \theta_1)\), \(\theta_2 = (C_2, r_2, \bar \theta_2)\), \(C_1 = D \lor A\), \(C_2
    = \lnot A \lor E\), and \(C = D \lor E\).  Inductively, we can transform
    \(\operatorname{annotate}(\theta_1, \sigma)\) into a new Metis theorem \(\theta_1' = (C_1',
    r_1', \bar \theta_1')\) with \(C_1' \subseteq C_1 \sigma\) and
    \(\operatorname{annotate}(\theta_2, \sigma)\) into \(\theta_2' = (C_2', r_2', \bar \theta_2')\)
    with \(C_2' \subseteq C_2 \sigma\).  Since instantiation can shorten clauses, we distinguish
    between two cases: In the first case, \(A \sigma\) is contained in the clauses \(C_1'\) and
    \(C_2'\), so we obtain \(D'\) and \(E'\), where \(C_1' = D' \lor A \sigma\) and \(C_2' = \lnot A
    \sigma \lor E'\).  Then we can construct the new Metis theorem \((C', r', \bar \theta') = (D'
    \lor E', \textsc{Resolve}, [\theta_1', \theta_2'])\) and \(C' \subseteq C \sigma\) holds since
    \(D' \lor E' \subseteq D \sigma \lor E \sigma = C \sigma\).  In the second case, we find
    \(\theta' = (C', r', \bar \theta') \in \{\theta_1', \theta_2'\}\) such that \(C'\) does not
    contain \(A \sigma\).  Then \(C' \subseteq C \sigma\) holds, and we can use \(\theta'\) as the
    new Metis theorem, thereby removing the resolution step as well as the other subproof.
\end{proof}
%\end{rep}

\maybesubsection{Translation of Metis Terms to Isabelle Terms}

Once the instantiations of the Metis clauses have been inferred, a \emph{translation procedure}
converts the contained Metis terms into Isabelle terms that can be presented to the user.

The list obtained from the \(\operatorname{infer}\) function is filtered to include only those pairs
\((C, \sigma)\) where the Metis clause~\(C\) is derived from an Isabelle fact~\(\varphi\), and not
from the negated goal.  The procedure then continues to translate only Metis terms~\(t\) of
substitution elements \((x \mapsto t) \in \sigma\) where the Metis variable~\(x\) occurs in \(C\)
and corresponds to a free variable in \(\varphi\).  There may be other Metis variables, including
those corresponding to type variables or quantified variables.  The direct instantiation of type
variables is seldom useful, since types can be inferred from the instantiated terms.  Metis
variables that correspond to quantified Isabelle variables may emerge during clausification;
however, since these are bound in Isabelle, there is no way to refer to them.

The translation procedure must decode constructs that are introduced by the translation from
polymorphic higher-order logic to untyped first-order logic in \textit{metis}.  This decoding uses
the same code as the proof reconstruction in \textit{metis} \cite{Paulson2007}, where Metis terms
emerge during reconstruction of the \textsc{Subst} inference rule.  The introduced constructs
include the encoding of Isabelle symbols using Metis constants, the encoding of free variables using
Metis variables, and the encoding of partial application using a distinguished binary symbol
\(\mathtt{app}\) \cite{Meng2007}.  For example, \(\mathtt{map}\;\mathtt{f}\) and
\(\mathtt{map}\;\mathtt{f}\;xs\) might be translated to \(\mathtt{app}(\mathtt{map},\mathtt{f})\)
and \(\mathtt{app}(\mathtt{app}(\mathtt{map},\mathtt{f}), xs)\), where \(\mathtt{map}\) and
\(\mathtt{f}\) are Metis constants and \(xs\) is a Metis variable.  While decoding these constructs
is straightforward, decoding the type information (for which there are several encodings
\cite{Blanchette2016}) was considered too complicated for proof reconstruction in \textit{metis}, so
type inference is used instead \cite{Paulson2007}.

The resulting Isabelle terms may still contain \emph{Skolem terms} and encodings of
\emph{\(\lambda\)-abstractions}.
Such \textit{metis}-internal constructs may not appear in terms suggested to
the user. Thus our translation procedure must eliminate them.

Skolem terms are eliminated by replacing them with a \emph{wildcard}, which is displayed as `\(\_\)'
and corresponds to a fresh free variable (as demonstrated in Example~\ref{ex:skolem-lambda}).
Consequently, the instantiated facts may still contain free variables, for which Metis must
substitute the corresponding Skolem terms.  It does not suffice to simply replace the Skolem symbols
with `\(\_\)'; their arguments must also be removed.  This is necessary because the arity of Skolem
symbols can change due to the instantiation of facts.  For example, the Skolem symbol introduced
for \(\exists y .\; x < y\) depends on \(x\) and therefore requires an argument, whereas the Skolem
symbol for the instance \(\exists y .\; 0 < y\) does not require any argument.

The \emph{metis} proof method encodes \(\lambda\)-abstractions
using either \(\mathtt{SKBCI}\) combinators \cite{Turner1979} or \(\lambda\)-lifted supercombinators
\cite{Hughes1982}. To eliminate these combinators, the translation
procedure replaces them with their definition, followed by a \(\beta\eta\)-reduction of the
resulting terms.  For example, the term \(\lambda x .\; 0\) is encoded as \(\mathtt{K}\;0\), where
\(\mathtt{K}\;a\;b = a\).  When the translation procedure detects the term
\(\mathtt{K}\;0\;1\), it replaces \(\mathtt{K}\) with \(\lambda a .\; \lambda b .\; a\) and
produces the term \(0\) after \(\beta\)-reduction.  If \(\lambda\)-abstractions remain
after \(\beta\eta\)-reduction, the bound variables may have different names than in the original
\(\lambda\)-abstractions (as demonstrated in Examples~\ref{ex:skolem-lambda} and~\ref{ex:ext}), but this is not a problem since Isabelle equates terms
up to \(\alpha\)-equivalence.

In the final step of the translation procedure, all remaining free variables (i.e., all free
variables occurring in the Isabelle terms and all uninstantiated variables in the clauses) are
instantiated with the Isabelle polymorphic constant \(\mathtt{undefined}\), which provides a witness
for every type's inhabitedness.  These variables emerge when the proof was possible without concrete
terms. For example, \(\mathtt{a} = \mathtt{b}\) can be derived from \(x + \mathtt{a} = x +
\mathtt{b}\) without choosing a value for \(x\).  Instantiation of such variables is sensible, since
it restricts the search space, but users can prevent this behavior by disabling the
\textit{metis\_instantiate\_undefined} option, in which case the variables are replaced by `\(\_\)'.

\maybesubsection{Instantiation of Isabelle Facts}

The translation procedure yields a list of Isabelle facts and instantiations of their free
variables. A single fact may possess multiple instantiations
if it is used multiple times in the proof.  To reduce the number of instantiations and to
avoid duplicates, we merge instantiations whenever possible:

\begin{definitionx}
    A (\emph{variable}) \emph{instantiation} \(\{x_1 \mapsto t_1, \dots, x_n \mapsto t_n\}\) for an
    Isabelle fact \(\varphi\) is a partial function from variable names to Isabelle terms in which
    all variable names \(x_1, \dots, x_n\) occur in \(\varphi\) as free variables.  Two
    instantiations \(\iota_1, \iota_2\) for the same fact \(\varphi\) can be \emph{merged} into
    \(\iota_1 \cup \iota_2\) if \(\iota_1(x)\) equals \(\iota_2(x)\) up to \(\alpha\)-equivalence
    for all shared variable names \(x \in \operatorname{dom}(\iota_1) \cap
    \operatorname{dom}(\iota_2)\).
\end{definitionx}

Merging can be used to avoid duplicates, since every instantiation can be merged with itself.
Conversely, if two instantiations stem from the same clause \(C\) and can be merged, they are equal
up to \(\alpha\)-equivalence.  This is because the translation procedure considers only variables
occurring in \(C\), and all uninstantiated variables of \(C\) are instantiated with
\(\mathtt{undefined}\) or `\(\_\)'.  As a result, the domains of the instantiations are exactly the
variable names occurring in \(C\) that correspond to a free variable in the fact.

By the above argument, if different instantiations can be merged, they must stem from different
clauses.  In practice, \textit{metis} will still succeed with the merged instantiation, since the
clausifier will split the instantiated fact into the instantiated clauses again.  Consequently,
merging does not lose information.

\begin{examplex}
    Consider the following Isabelle fact \(\varphi\):
    \[\operatorname{even}\;n \longrightarrow 0 \leq x ^ n \land (-y) ^ n = y ^ n\]
    The clausifier splits it into two clauses.  Suppose that both clauses are used in a
    \textit{metis} proof and we inferred the instantiations \(\iota_1 = \{x \mapsto \mathtt{a}{,}\;
    n \mapsto 2\}\) and \(\iota_2 = \{y \mapsto\nobreak \mathtt{b}{,}\; n \mapsto 2\}\).  Since the
    variable name \(n\) is mapped to the same term in both instantiations, these can be merged to
    \(\iota_1 \cup \iota_2 = \{x \mapsto \mathtt{a}{,}\; y \mapsto \mathtt{b}{,}\; n \mapsto 2\}\).
    By instantiating \(\varphi\) with \(\iota_1 \cup \iota_2\), we obtain the following new fact:
    \[\operatorname{even}\;2 \longrightarrow 0 \leq \mathtt{a} ^ 2 \land (-\mathtt{b}) ^ 2 =
    \mathtt{b} ^ 2\]
    If \textit{metis} is invoked again using this fact instead of the original fact, the clausifier
    will again split it into the two clauses, instantiated with \(\iota_1\) and \(\iota_2\),
    respectively.
\end{examplex}

Since we use type inference instead of reconstructing type information,
the translation
procedure may produce Isabelle terms with overly generic types.  Given that types can be displayed in
Isabelle (e.g., using the \textit{show\_types} option) and overly generic types can prevent
the folding of abbreviations, we attempt to concretize the types of the terms of a variable
instantiation \(\iota\) for a fact \(\varphi\).  We achieve this through type unification
\cite{Nipkow1995}: The types of the terms of \(\iota\) are unified with the types of the
corresponding variables in \(\varphi\), and the unifier is then applied to all the types contained
in the terms of \(\iota\).  It is crucial to find a single unifier for the entire instantiation
\(\iota\), since the same type variable may appear in multiple types of the free variables in
\(\varphi\), and they must be synchronized.

Finally, if instantiations have been found, \textit{metis} suggests that the user replaces the current
\textit{metis} call with a new call using the instantiated facts.  Each instantiated fact is
displayed by its name together with an annotation that
specifies to instantiate it
with the terms of the corresponding instantiation.
If there are multiple instantiations of a
single fact, the fact is displayed multiple times.  By replacing the call, the user may
get a more readable and faster proof.

\section{Sledgehammer Extension}
\label{sec:sh-extension}

\looseness=-1
Since \textit{metis} calls are usually generated by Sledgehammer, we also extend
Sledgehammer's {proof reconstruction module} to enable direct generation of proofs with
instantiated facts.  By using Sledgehammer's {preplay module}, this extension
can not only generate faster and more readable proofs
but also transform timeouts
into successes as well as produce simpler proofs using
other proof methods than \textit{metis}.

\maybesubsection{Preplay with Instantiations}

Sledgehammer's proof reconstruction is started after an external ATP has found and minimized a
proof.  The facts referenced in the ATP proof are then used to generate an Isabelle proof.  The
\textit{metis} extension we presented in Section~\ref{sec:metis-extension} can now be invoked to
infer the instantiations of these facts.  To do so, \textit{metis} must first succeed---both the
Metis proof (for inferring the instantiations) and the reconstruction in Isabelle (for ensuring that
the instantiations are type-correct, if an unsound type encoding was used) are essential.

After instantiating the facts, Sledgehammer's preplay tries out the proof with the instantiated
facts.  Even though \textit{metis} has already found a proof, this is beneficial for a number of
reasons:
\begin{enumerate}
    \item
        The preplay module includes an additional minimization tool that repeatedly invokes a
        successful proof method with subsets of the facts.  Its purpose is to reduce the number of
        facts and thereby simplify the generated Isabelle proof.
        Instantiated facts might be redundant and, consequently, can be eliminated by
        this minimization.
        Consider the goal \(\mathtt{a} < -\mathtt{b} \longleftrightarrow \mathtt{b} < -\mathtt{a}\)
        and the fact \(x < -y \longleftrightarrow y < -x\).  This fact is
        translated to two Metis clauses, corresponding to the two implications \(x < -y
        \longrightarrow y < -x\) and \(y < -x \longrightarrow x < -y\).  Since both clauses are
        equal up to the naming of variables, Metis uses only the first one to prove both directions
        of the goal.  Accordingly, the instantiations \(\{x \mapsto \mathtt{a}{,}\; y \mapsto
        \mathtt{b}\}\) and \(\{x \mapsto \mathtt{b}{,}\; y \mapsto \mathtt{a}\}\) are inferred,
        resulting in two instantiated facts.  Since only one of these is necessary, the minimization
        tool removes one of them.%
    \smallskip
    \item
        Via a mechanism called \emph{try0}, preplay tries a variety of proof methods \cite{blanchette-et-al-2016-isar}, including multiple
        \textit{metis} calls with different options (describing the encoding of types and
        \(\lambda\)-abstractions) and other standard proof methods, such as \textit{simp}
        \cite{Nipkow1989}, \textit{blast} \cite{Paulson1999}, and \textit{auto} \cite{Paulson1994}.
        This can result in even faster Isabelle proofs, since other proof methods are frequently
        faster than \textit{metis} but sometimes require instantiated facts to find any proof at all
        (as demonstrated in Example~\ref{ex:speedup-auto}).  Additionally, Isabelle proofs resulting
        from methods such as \textit{simp}, \textit{blast}, and \textit{auto} are often simpler and
        more readable, since they do not require the user to provide all
        the necessary facts for the proof.
        Minimization removes the facts that are irrelevant for these proof methods.%
    \smallskip
    \item
        According to Theorem~\ref{thm:instantiated-axiom-clauses}, a new Metis proof derived from
        the instantiated Metis clauses is possible.  However, this does not necessarily extend to
        Isabelle proofs using \textit{metis}.  A new \textit{metis} proof using the instantiated
        Isabelle facts may not be possible, since their encoding is not guaranteed to
        produce the same Metis clauses again.  This is mostly due to \(\beta\eta\)-conversion (as
        demonstrated in Example~\ref{ex:ext}).  This rarely happens in practice but
        is nonetheless one reason to preplay the new \textit{metis} proof.
\end{enumerate}

During preplay, each proof method is executed for a predetermined duration (called the \emph{preplay
time limit}, which defaults to \(1\)\,s).  The instantiation of facts often speeds up proofs,
possibly turning timeouts into successes.  However, in order to instantiate the facts,
\textit{metis} must first succeed.  Consequently, the \textit{metis} call used to infer the
instantiations is given more time than the proof methods during preplay.  More precisely, it is executed
for five times the preplay time limit (i.e., \(5\)\,s by default).

An alternative would be to increase the preplay time limit in the first place.
However, this would work less well in practice, because users typically insert
the generated proofs directly into the Isabelle proof text. Given that
these are re-executed each time the text is processed, multiple slow proofs
could substantially increase the overall processing time.

Our Sledgehammer extension
can be controlled through the three-valued Sledgehammer option \textit{instantiate}.  If this option
is set to \textit{false}, the extension is disabled.
If the option is set to \textit{true}, \textit{metis} is started with the
\textit{metis\_instantiate} option enabled directly after an ATP has found and minimized a proof, at
the beginning of Sledgehammer's proof reconstruction module.  In most cases, this leads to Isabelle
proofs with instantiated facts; otherwise, a proof without instantiations is displayed.  The latter
case may occur if \textit{metis} is unable to find a proof (but, e.g., \textit{auto} is able to),
the proof does not use any facts, or there are no instantiations (e.g., if the facts contain no free
variables).

Finally, if the option is set to \textit{smart}, the instantiation of facts is started only if
preplay failed (i.e., no proof method was successful within the preplay time limit).  If the
instantiation was successful, preplay is invoked once more with the instantiated facts, as usual.
Therefore, Isabelle proofs with instantiated facts are displayed only if \textit{metis} takes more
time than the preplay time limit to prove the goal from the original facts and instantiation can
speed up the proof so that it takes less time than the preplay time limit (as demonstrated in
Example~\ref{ex:timeout}).

Although setting the option to \textit{true} leads to faster, simpler, and more readable proofs, it
is too disruptive.  Most users, most of the time,
are satisfied with uninstantiated facts, which are less verbose.  They are
pleased if they occasionally get a proof with instantiated facts that would have failed otherwise.
Therefore, we make \textit{smart} the default.

\maybesubsection{Extensionality for Metis}

Extensionality states that two functions are equal if they yield the same
results for the same arguments.  In Isabelle, extensionality is a basic axiom called
\textit{ext}:
\[(\forall x .\; f\;x = g\;x) \longrightarrow f = g\]
Since Metis targets first-order logic, it is unaware of this principle.
Instead, \textit{ext} must be passed to \textit{metis}
(which uses a first-order encoding) when extensionality is required.
Since this enlarges the search space, this is not done
by default.

However, the need for extensionality can change as facts are instantiated.
In other words, \textit{ext} may not be necessary for a proof derived from the original facts,
but it may be necessary for a proof derived from the instantiated facts, and vice versa.  This
phenomenon is mostly due to Isabelle's application of \(\beta\eta\)-conversion.

\begin{examplex}
\label{ex:ext}
    Recall the fact \(\mathtt{surj}\;f \longrightarrow (\exists x .\; f\;x = y)\), and suppose that we want to prove the following goal:
    \[\mathtt{surj}\;(\lambda x .\; \lambda y .\; \mathtt{g}\;y\;x) \longrightarrow (\forall x
    .\; \mathtt{P}\;(\lambda y .\; \mathtt{g}\;y\;x)) \longrightarrow \mathtt{P}\;\mathtt{h}\]
    To encode the \(\lambda\)-abstractions, we use \(\lambda\)-lifting \cite{Hughes1982}
    and thus introduce a new {supercombinator} \(\mathtt{A}\) with the definition
    \(\mathtt{A}\;a\;b = \mathtt{g}\;b\;a\), leading to a new goal:
    \[\mathtt{surj}\;\mathtt{A} \longrightarrow (\forall x .\; \mathtt{P}\;(\mathtt{A}\;x))
    \longrightarrow \mathtt{P}\;\mathtt{h}\]
    We invoke \textit{metis}, which infers that \(f\) should be instantiated with \(\mathtt{A}\) and
    \(y\) should be instantiated with \(\mathtt{h}\).
    As described in
    Section~\ref{sec:metis-extension}, \(\mathtt{A}\) is replaced by \(\lambda a .\; \lambda b .\;
    \mathtt{g}\;a\;b\) again, resulting in the following instantiated fact:
    \[\mathtt{surj}\;(\lambda a .\; \lambda b .\; \mathtt{g}\;b\;a) \longrightarrow (\exists x
    .\; (\lambda b .\; \mathtt{g}\;b\;x) = \mathtt{h})\]
    Notice that Isabelle
    applied \(\beta\)-reduction, resulting in the modified \(\lambda\)-abstraction \(\lambda b .\;
    \mathtt{g}\;b\;x\).  If we attempt a proof with \textit{metis} and \(\lambda\)-lifting again,
    this \(\lambda\)-abstraction will be encoded as another supercombinator \(\mathtt{B}\) with the
    definition \(\mathtt{B}\;a = \mathtt{g}\;a\;\mathtt{sk}\), where \(\mathtt{sk}\) is the Skolem
    constant introduced for the existentially quantified variable~\(x\).  Using \(\mathtt{A}\)'s
    and \(\mathtt{B}\)'s definitions, Metis
    can prove that \(\mathtt{A}\;\mathtt{sk}\;z = \mathtt{B}\;z\) for every~\(z\), but extensionality
    is needed to conclude that \(\mathtt{A}\;\mathtt{sk} = \mathtt{B}\) and thus to exchange the two
    terms and prove the goal.  Therefore, \textit{metis} can prove the goal from the
    instantiated fact only if we add the fact \textit{ext}.
\end{examplex}

Thus, \textit{ext} may be necessary for a proof derived
from the instantiated facts.  Conversely, if \(\beta\eta\)-conversion eliminates
\(\lambda\)-abstractions, \textit{ext} might be unnecessary for a proof derived from the
instantiated facts even though it is needed for a proof derived from the original facts.  Therefore,
we also extend the preplay module so that a few \textit{metis} calls with the
additional fact \textit{ext} are tried after other proof methods have failed, including multiple
\textit{metis} calls with different options. These new \textit{metis} calls use \(\lambda\)-lifting
as the encoding of \(\lambda\)-abstractions, since \(\lambda\)-lifting benefits
more from \textit{ext} than the \(\mathtt{SKBCI}\) combinators \cite{Meng2007}.

The \textit{metis} calls with the fact \textit{ext} are also useful regardless of instantiations,
since some ATPs, such as Zipperposition \cite{vukmirovic-et-al-2021-making}, are based on
higher-order logic and already include extensionality.  As a result, their proofs do not need
\textit{ext}, and proof reconstruction might be unsuccessful without the new \textit{metis} calls.

\section{Evaluation}
\label{sec:evaluation}

Our empirical evaluation is based on the repository revision d3c0734059ee
(October 25, 2024) of Isabelle and 4082096ade5a (October 25, 2024) of the
\emph{Archive of Formal Proofs} \cite{Afp2015}.

\maybesubsection{Setup}

We performed the evaluation using the Slurm batch system \cite{Jette2023} on the computer resources
of the Institute for Informatics at Ludwig-Maximilians-Universität München.  We
requested \(32\)\,GiB of RAM and 8 logical processors (CPU threads).
Given that not all Sledgehammer steps and ATPs are deterministic and time limits are used, the results
are not entirely reproducible. To increase reproducibility, we used a fresh Isabelle installation
and reset the state of the learning-based relevance filter MaSh
\cite{blanchette-et-al-2016-mash} before each evaluation run.

We used the testing and evaluation tool Mirabelle \cite{Desharnais2022}, which is included in
Isabelle.  Mirabelle applies a selected \emph{action} to each selected \emph{goal}.
We used the same \(50\) \emph{Archive} entries as Desharnais et al.~\cite{Desharnais2022}, from which we randomly
selected \(100\)~goals per entry, resulting in a total of \(5\,000\) goals.  As the action, we
invoked Sledgehammer with several options:
\begin{itemize}
    \item
        The \textit{try0} option was disabled so that Sledgehammer's preplay
        exclusively tried \textit{metis}.  This was done to test \textit{metis} extensively, since
        we extended this proof method, and to keep the evaluation feasible in reasonable time, since other
        proof methods frequently caused the Isabelle process to abort for
        technical reasons.%
    \smallskip
    \item
        The \textit{provers} option for specifying the external ATPs was set to use the high-performance~\cite{Desharnais2022}
        superposition provers E% \cite{vukmirovic-et-al-2023}
        , Vampire% \cite{bhayat-suda-2024}
        , and
        Zipperposition% \cite{vukmirovic-et-al-2021-making}
        . We excluded satisfiability-modulo-theories (SMT) solvers since these support theories, such
        as linear arithmetic, that are not built into~\textit{metis}.%
    \smallskip
    \item
        The \textit{strict} option was enabled to force the use of sound type
        encodings for the external ATPs to produce type-correct proofs that should
        be reconstructable.
\end{itemize}

We performed two evaluation runs, with Sledgehammer's \textit{instantiate} option set to
\textit{smart} in the first run and to \textit{true} in the second run.
From the Mirabelle output of each run, we extracted the number of successful Isabelle
proofs with and without instantiations, the average execution times of Sledgehammer and
\textit{metis}, and the kinds of \textit{metis} calls used in the generated Isabelle proofs.  With
this information, we try to answer the following research questions:
\begin{enumerate}
    \item% 1
        Is it possible to reconstruct additional proofs by instantiating the
        facts, and in how many cases is a proof derived from the instantiated facts
        impossible?%
    \smallskip
    \item% 2
        What are the advantages of enabling Sledgehammer's \textit{instantiate} option?
        That is, how often are Isabelle proofs with instantiated facts actually
        presented to users, how much longer do Sledgehammer executions take, and how
        much faster are Isabelle proofs?%
    \smallskip
    \item% 3
        What do Sledgehammer-generated \textit{metis} calls look like after instantiating the facts,
        and what are the benefits of adding the fact \textit{ext}?
\end{enumerate}

\maybesubsection{Results}

We consider only Sledgehammer invocations where an external ATP has found a proof.  Sledgehammer's
proof reconstruction module initially attempts to generate a \emph{one-line proof} consisting of a
single proof method invoked with facts, which can now have instantiations thanks to our extensions.
If this attempt is unsuccessful, Sledgehammer tries to transform the ATP proof into an {Isar} proof
\cite{blanchette-et-al-2016-isar}.  If this also fails, the proof reconstruction is considered to
have \emph{failed}.  The number of occurrences of each of these outcomes in each evaluation run is
shown in Table~\ref{tab:result-reconstruction}.

\begin{table}
    \caption{Proof reconstruction for Sledgehammer invocations yielding ATP proofs}
    \label{tab:result-reconstruction}
    \centering
    \begin{tabular}{lrrrrr}
        \toprule
        & \multicolumn{2}{c}{One-line proofs} \\
        \cmidrule{2-3}
        Evaluation run & W/o inst. & With inst. & Isar proofs & Failed & Total \\
        \midrule
        1 (\textit{instantiate} = \textit{smart}) &
            \(3\,128\) & \(44\) & \(37\) & \(164\) & \(3\,373\) \\
        2 (\textit{instantiate} = \textit{true}) &
            \(805\) & \(2\,349\) & \(45\) & \(159\) & \(3\,358\) \\
        \bottomrule
    \end{tabular}
\end{table}

In the first evaluation run, there were a total of \(3\,373\) Sledgehammer invocations yielding ATP
proofs.  Whereas \(3\,128\) ATP proofs (\(92.7\%\)) were reconstructed directly as one-line proofs
without instantiations, \(44\) additional ATP proofs (\(1.3\%\)) could be reconstructed by
instantiating the facts and thereby speeding up the \textit{metis} proofs.  This number may seem
low, but this corresponds to \(18.0\%\) of cases where there was no one-line proof before.  The
remaining \(6.0\%\) of ATP proofs could not be reconstructed as one-line proofs.  This number may
seem alarmingly high, but detailed Isar proofs were available in \(1.1\%\) of the cases, and in
practice the \textit{try0} option is usually enabled, which leads to further proof methods being
tried in addition to \textit{metis}.  We note that our approach brings similar benefits to
Sledgehammer's success rate as Isar proof reconstruction.

In the second evaluation run, there were a total of \(3\,358\) Sledgehammer invocations with ATP
proofs.  In \(2\,349\) cases (\(70.0\%\)), Sledgehammer suggested a one-line proof with instantiated
facts.  In \(805\) cases (\(24.0\%\)), Sledgehammer suggested a one-line proof without instantiated
facts. The raw evaluation data reveals that
in most of these cases, the proof did not use any facts or there were no instantiations.

However, sometimes the instantiations were successfully inferred, but preplay failed to generate a
\textit{metis} proof derived from the instantiated facts.  According to the raw data, there were
\(5\) such cases (\(0.1\%\)) in the first evaluation run and \(18\) (\(0.5\%\)) in the second.
There are several reasons for this, including the following:
\begin{itemize}
    \item
        The instantiations did not speed up the \textit{metis} proof enough to take less time than
        the preplay time limit.  The instantiations could have even prolonged the
        proof, since they modify the problem and could affect Metis's heuristics.%
    \smallskip
    \item
        Isabelle's extremely flexible syntax allows ambiguities and does not guarantee that all
        terms can be parsed back \cite{blanchette-et-al-2016-isar}.
\end{itemize}

\begin{table}
    \caption{Average execution times for a one-line proof}
    \label{tab:result-times}
    \centering
    \begin{tabular}{lrr}
        \toprule
        Evaluation run & Sledgehammer & Generated \textit{metis} proof \\
        \midrule
        1 (\textit{instantiate} = \textit{smart}) & \(34\,185\)\,ms & \(221\)\,ms \\
        2 (\textit{instantiate} = \textit{true}) & \(34\,497\)\,ms & \(137\)\,ms \\
        \bottomrule
    \end{tabular}
\end{table}

\looseness=-1
Table~\ref{tab:result-times} shows the average execution times of Sledgehammer invocations that
successfully generated a one-line \textit{metis} proof.  The average execution time of Sledgehammer
increased by \(312\)\,ms (\(+0.9\%\)) from the first to the second evaluation run, which can be
explained by the additional time required for instantiating facts.  However, this did not have a
substantial impact on the overall duration of an average Sledgehammer execution.  By contrast, the
average execution time of \textit{metis} decreased by \(84\)\,ms (\(-38.0\%\)), indicating that the
instantiation of facts markedly speeds up proofs.  Notice that in about \(25\%\) of cases (\(805\)
and \(44\) in Table~\ref{tab:result-reconstruction}), the \textit{metis} calls generated in the two
evaluation runs were both without or both with instantiations. Consequently, for the remaining
\(75\%\) of cases, it can be deduced that the average execution time decreases by more than
\(38\%\).  If those proofs are inserted into an Isabelle proof text, this improvement applies
every time the text is processed. In particular, every one of the frequent
regression tests of the \emph{Archive of Formal Proofs} with its more than \(90\,000\)
\emph{metis} calls could benefit.

In Sledgehammer's preplay, a series of \textit{metis} calls are tried, each involving different
options that describe the encoding of types and \(\lambda\)-abstractions.  Initially, the default
options (type encoding \(\mathsf{a}\) \cite{Blanchette2016} and \(\mathtt{SKBCI}\) combinators
\cite{Turner1979}) are tested, resulting in a standard \textit{metis} call.  If this fails,
Sledgehammer tries multiple alternative options, selecting the fastest \textit{metis} call with
options.  If this also fails, Sledgehammer tries the new \textit{metis} calls with the additional
fact \textit{ext}.  The number of occurrences of each of these kinds of \textit{metis} calls in
one-line proofs in each evaluation run is shown in Table~\ref{tab:result-calls}.

\begin{table}
    \caption{Kinds of \textit{metis} calls in one-line proofs}
    \label{tab:result-calls}
    \centering
    \begin{tabular}{lrrrr}
        \toprule
        Evaluation run & Standard \textit{metis} & With options & With \textit{ext} & Total \\
        \midrule
        1 (\textit{instantiate} = \textit{smart}) & \(2\,801\) & \(318\) & \(53\) & \(3\,172\) \\
        2 (\textit{instantiate} = \textit{true}) & \(3\,008\) & \(110\) & \(36\) & \(3\,154\) \\
        \bottomrule
    \end{tabular}
\end{table}

The number of standard \textit{metis} calls increased from the first evaluation run (\(88.3\%\)) to
the second (\(95.4\%\)).  This observation suggests that Metis's proof search becomes more
efficient after the instantiation of facts, since special encodings are needed less often.
Moreover, adding \textit{ext} was effective, since it was used in \(1.7\%\) and \(1.1\%\) of the
proofs, respectively, which would have failed otherwise.

Regarding our research question 1, the results show that Sledgehammer can reconstruct additional
one-line proofs by instantiating the facts, and that a proof
derived from the instantiated facts is almost always possible. Regarding question 2, we observe that enabling
Sledgehammer's \textit{instantiate} option leads to proofs with instantiations in \(70\%\) of the
cases, which may help users understand the generated proofs.  Sledgehammer executions also take a
bit longer, but the execution time of the generated Isabelle proofs is substantially reduced, which
is much more important.
Regarding question 3, we see that there are more standard \textit{metis} calls without options after
instantiating the facts and that adding the fact \textit{ext} brings similar benefits to
Sledgehammer's success rate as instantiating facts.

\section{Related Work}

The Metis ATP \cite{Hurd2003}, on which our work is based, was
developed by Hurd with the goal of integrating it into the HOL proof assistant
\cite{Gordon91}.
Metis was integrated in Isabelle by Paulson and Susanto \cite{Paulson2007}.
Metis's fine-grained proofs were particularly useful to us.
Beyond the HOL and Isabelle integrations, there is also a Metis proof checker
in Agda developed by Prieto-Cubides and Sicard-Ram\'\i{}rez \cite{AgdaMetis}.

Our tool joins the ranks of various proof analysis tools. The GAPT (General
Architecture for Proof Theory) framework by Ebner et
al.~\cite{DBLP:conf/cade/EbnerHRRWZ16} consists of data
structures, algorithms, parsers, and more, with the aim of supporting proof theory
applications and ATPs. Other proof-manipulating frameworks are ProofCert by
Miller and colleagues
\cite{DBLP:conf/lfmtp/Miller13,DBLP:journals/jar/ChihaniMR17} and Dedukti by
Dowek and
colleagues~\cite{DBLP:journals/corr/abs-2311-07185,DBLP:conf/fossacs/BlotDTW24}.

Our work is also loosely related to the experimental reconstruction of ATP
proofs as detailed Isabelle proofs, or Isar proofs \cite{Wenzel99}, by Blanchette et
al.~\cite{blanchette-et-al-2016-isar}.

\section{Conclusion}

We extended Isabelle's \emph{metis} proof method and the Sledgehammer tool to
infer variable instantiations from proofs and present them to users. This
speeds up Sledgehammer's proof reconstruction and increases its success rate. It
also helps users to understand the proof without inspecting all of its details.

We see three main directions for future work. First, we could provide type
annotations when parsing of the instantiations is ambiguous. Second,
we could provide instantiations not only for one-line proofs but also for
individual steps in detailed Isar proofs. Third, we could extend the approach
to more ATPs in order to obtain variable instantiations even when \emph{metis} without
instantiations fails. The superposition provers E% \cite{vukmirovic-et-al-2023}
, SPASS% \cite{blanchette-et-al-2012-spass}
, Vampire% \cite{bhayat-suda-2024}
, and Zipperposition %\cite{vukmirovic-et-al-2021-making}
are difficult to integrate,
because they fail to preserve the connection between the variables before and
after clausification. The \emph{metis} proof method's clausifier, by contrast,
preserves variable names. As for SMT solvers, some of them, including cvc5% \cite{barbosa-et-al-2022}
, can be asked to output the instantiations they needed to show
unsatisfiability. Then there is no need to analyze the proofs to infer the
instantiations.

\subsubsection{Acknowledgments.}

\small
We thank Xavier G\'en\'ereux, Elisabeth Lempa, Jannis Limperg, Kirstin Peters,
Mark Summerfield, and the anonymous reviewers for suggesting various textual
improvements.
Blanchette's research was cofunded by the European Union (ERC, Nekoka,
101083038). Views and opinions expressed are however those of the authors only
and do not necessarily reflect those of the European Union or the European
Research Council. Neither the European Union nor the granting authority can be
held responsible for them.

\subsubsection{Disclosure of Interests.}

\small
The authors have no competing interests to declare that are relevant to the
content of this paper.

\bibliographystyle{splncs04}
\bibliography{bib}

\end{document}